\documentclass[9pt,a4paper,twocolumn]{article}
\pdfoutput = 1

\usepackage{color}
\usepackage{dcolumn}
\usepackage{bm}
\usepackage{soul}
\usepackage{braket}
\usepackage[T1]{fontenc} 
\usepackage{float}
\usepackage{amsmath,amsfonts,amssymb,eqnarray}  
\usepackage{amsthm, color, soul}
\usepackage{graphicx}  
\usepackage{array,booktabs}
\usepackage[pdftex,hypertexnames=false]{hyperref} 
\usepackage[affil-it]{authblk}
\usepackage[font={small}]{caption}
\usepackage{cite}
\usepackage[left=2cm,right=2cm,top=1.6cm]{geometry}

\title{\textbf{Air-core fiber distribution of hybrid vector vortex-polarization entangled states}} 
\author{Daniele Cozzolino$^1$, Emanuele Polino$^2$, Mauro Valeri$^2$, Gonzalo Carvacho$^2$, Davide Bacco$^1$, Nicol\`o Spagnolo$^2$, Leif Katsuo Oxenl\o we$^1$, Fabio Sciarrino$^{2,*}$}

\affil{$^1$ \small CoE SPOC, Dep. Photonics Eng., Technical University of Denmark, Kgs. Lyngby 2800, Denmark}
\affil{$^2$ Dipartimento di Fisica, Sapienza Universit\`{a} di Roma, Piazzale Aldo Moro 5, I-00185 Roma, Italy}
\affil{Email: *fabio.sciarrino@uniroma1.it} 
\date{\vspace{-1em} \small   Dated: \today } 

\begin{document}
\pagestyle{plain}
\setcounter{page}{1}
\twocolumn[ 
\begin{@twocolumnfalse}
\maketitle
     \vspace{-0.8cm}
  \begin{abstract}
      \normalsize
         \vspace{1pt}
\noindent Entanglement distribution between distant parties is one of the most important and challenging tasks in quantum communication. Distribution of photonic entangled states using optical fiber links is a fundamental building block towards quantum networks.  Among the different degrees of freedom, orbital angular momentum (OAM) is one of the most promising due to its natural capability to encode high dimensional quantum states. In this article, we  experimentally demonstrate  fiber distribution of hybrid polarization-vector vortex entangled photon pairs. To this end, we exploit a recently developed air-core fiber which supports OAM modes. High fidelity distribution of the entangled states is demonstrated by performing quantum state tomography  in the polarization-OAM Hilbert space  after fiber propagation, and by violations of Bell inequalities and multipartite entanglement tests. The present results open new scenarios for quantum applications where correlated complex states can be transmitted by exploiting the vectorial nature of light.
\end{abstract}
  \end{@twocolumnfalse}]

\subsection*{Introduction}
Quantum communication requires the reliable transmission of quantized information carriers (qubits) among several and spatially separated parties~\cite{GisinQC}, towards development of quantum networks. In particular, protocols based on genuine quantum schemes like entanglement swapping~\cite{pan_swap,swap2,swap3,swap4}, superdense coding~\cite{kwiat_dense,superdense,superdense2,superdense3} and quantum teleportation~\cite{Bouw97,Bosc98,Lomb02,pan_tele,teleport2,teleport3,teleport4}, have to be adopted in future networks to access communication advantages that would be unattainable by using any classical resource. The key element of these schemes is  entanglement, which is one of the most distinctive quantum phenomena, predicted by Einstein, Podolsky and Rosen~\cite{EPR}, that defies the classical notion of local causality \cite{Bell}. Quantum correlations are an essential ingredient for quantum foundations studies and for different quantum information processes \cite{Horodecki,Nielsen,Brunner,Adesso}. Great interests has been devoted to the coherent distribution through optical fibers of such quantum correlations, since it constitutes the cornerstone for the future quantum networks \cite{Yin2017,Ursin2007,Honjo,timebin300km}. Several photon degrees of freedom can be employed for this task, such as frequency, orbital angular momentum (OAM), time and polarization~\cite{Honjo,timebin300km,Hubel,Poppe,Flamini19rev}. 
\begin{figure*}[ht!]
   \centering
    {\includegraphics[width=1\linewidth]{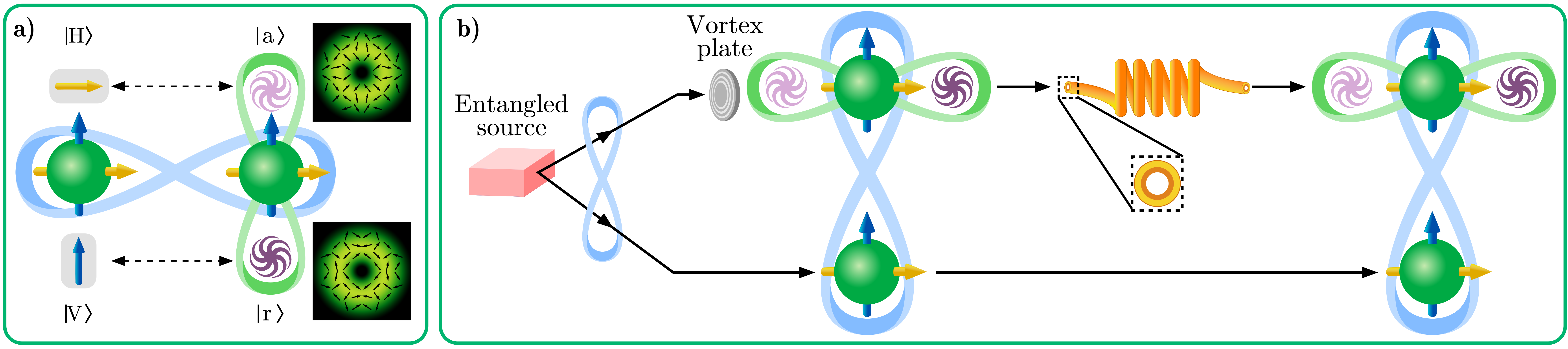}}
    \caption{\textbf{Hybrid entangled state transmission.} \textbf{a)} Hybrid VV-polarization entangled photon pair generated in the experiment:  entanglement in polarization of the photon pair (blue ribbon) and  entanglement between polarization and OAM of the single photon (green ribbon, VV state) are sketched. The inhomogeneous polarization patterns of the VV state $\ket{r}$ (bottom) and $\ket{a}$ (up) are explicitly shown. \textbf{b)}
    Schematic of the experiment: hybrid VV-polarization entangled state is generated by an initial polarization entangled photon pair. One photon of the pair encodes the VV state by the action of a vortex plate. The VV beam is transmitted through the air-core fiber. Finally,  state detection shows that hybrid VV-polarization entanglement (blue and green ribbons) is preserved after fiber transmission.}
\label{fig:cnpt}
\end{figure*}
In particular OAM of light is one of the most promising, albeit challenging to manipulate. Photons owning a non-zero OAM are characterized by the azimuthal phase dependence $e^{i\ell\phi}$, where $\ell\hbar$ is the amount of OAM carried by each photon, and $\ell$ is an unbounded integer value representing discrete quantum states~\cite{Allen1992,Molina2007}. Due to its unbounded nature, OAM has been largely investigated both for classical and quantum communications, being capable  of encoding high dimensional quantum states (qudits), which enhance the photon information capacity~\cite{Bozinovic2013,Wang2012,Willner2015,Kasper,Malik2016,Bavaresco2018,Erhard2018,Taira2018,Dada2011,krenn2014communication,bouchardunderwater,cinaunderwater}. Experimental investigations on OAM supporting fibers for classical communications have been reported \cite{Bozinovic2013,Wang2012,oam36fiber,supermode,ndagano2015fiber}. Still, distribution of quantum states through multimode fibers supporting OAM modes is a newborn research field, where only few experiments have been realized to date~\cite{Cozzolino2018,Sit2018}.
In reference~\cite{Cozzolino2018}, high-dimensional quantum states, encoded in weak coherent pulses, have been transmitted and detected through a 1.2 km length air-core fiber at telecom wavelength, demonstrating the feasibility of high-dimensional quantum communication and quantum key distribution protocols. In reference~\cite{Sit2018}, a solid core vortex fiber has been exploited, demonstrating the possibility of two-dimensional quantum communication with structured photons. Nonetheless, experiments investigating the transmission of entangled photon pairs still need to be explored.\\
In this work, we demonstrate distribution of hybrid entanglement between a linearly polarized photon and a vector vortex (VV) beam, \textit{i.e.} a doughnut-shaped beam with an inhomogeneous polarization pattern, at telecom wavelength. The VV beam is transmitted through a 5m long air-core fiber~\cite{Gregg2015}, whose very low mode mixing preserves OAM states and, in turn, hybrid entanglement. 
\begin{figure}[h!]
   \centering
    {\includegraphics[width=.98\linewidth]{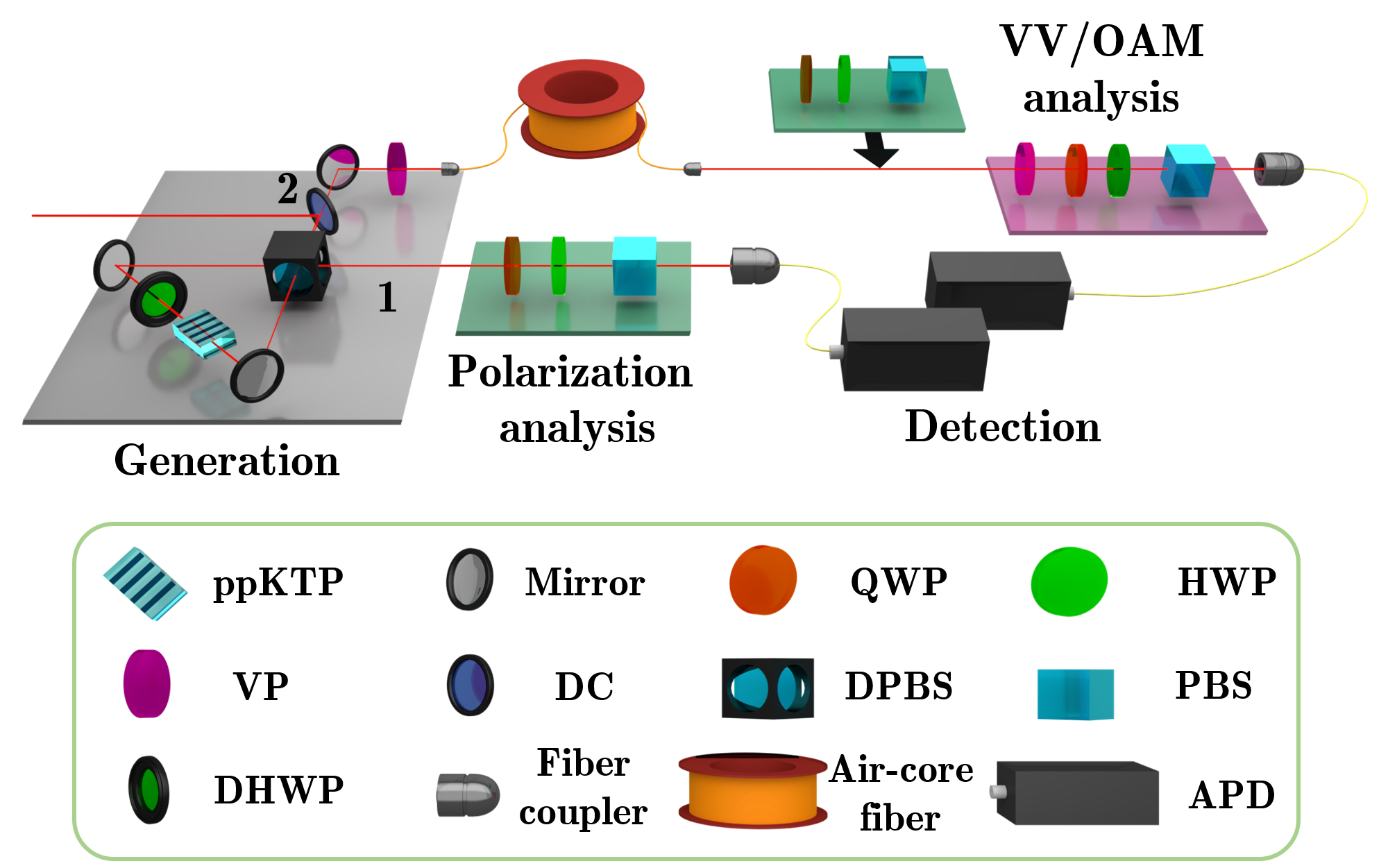}}
    \caption{\textbf{Experimental apparatus for the generation, distribution and analysis of the hybrid entangled states.} 
    Pairs of telecom polarization entangled photons are generated by exploiting a periodically poled titanyl phosphate crystal (ppKTP) in a Sagnac interferometer, which contains a dual-wavelength polarizing beam splitter (DPBS) and a dual half-wave plate (DHWP). Photons exiting along mode 1 are sent to a polarization analysis stage, composed of a quarter-wave plate (QWP), half-wave plate (HWP) and a polarization beam splitter (PBS). Photons along mode 2 pass through a dichroic mirror (DC), which separates the pump from the photons. Photons in mode 2 impinge on a vortex plate (VP) to generate a VV beam state and, in turn, the desired hybrid entangled state. The VV states are coupled to an air-core fiber and then measured with an OAM-polarization analysis stage composed of a second VP followed by a polarization analysis setup. To perform the measurements on the  polarization and OAM degrees of freedom independently, an additional polarization measurement stage has to be inserted before the OAM-to-Gaussian conversion regulated by the second VP. Finally, both photons are coupled into single-mode fibers linked to avalanche photodiode single photon detectors (APDs).
    }
\label{fig:setup}
\end{figure}
This peculiar feature opens up new scenarios and opportunities in quantum communications towards fiber based quantum networks, enabling the capability to employ high dimensional quantum states, embedded in the photon polarization and OAM degrees of freedom. 

\subsection*{Vector vortex beam and hybrid entanglement generation}
Vector vortex beams constitute a special class of vector beams, which are characterized by an inhomogeneous polarization distribution over their transverse profile \cite{VB}. In particular, a VV beam has an azimuthally varying polarization pattern, surrounding a central optical singularity \cite{ZhanVV,Cardano,Milione}. Due to their distinctive polarization distributions, VV beams  have shown unique features, making them appealing for different research purposes, \textit{e.g.} microscopy \cite{VVmicro}, optical trapping \cite{VVtrap,VVtrap2}, metrology \cite{VVmetro1,VVmetro2}, nanophotonics \cite{Buse18} and communication \cite{VVcomm1,VVcomm2,VVcomm3,VVcomm4,Damb12,Vall14,Carv17,Bouc18}. Formally, the state $\ket{R,\ell}$ ($\ket{L,\ell}$) describes a photon with uniform right (left) circular polarization carrying $\ell\hbar$ of OAM, and a VV beam can be conveniently described by a non-separable superposition of polarization-OAM eigenmodes. In this superposition, OAM quanta carried by the photons define the order $m$ of the VV beam.
\begin{figure*}[ht!]
   \centering
    {\includegraphics[width=1\linewidth]{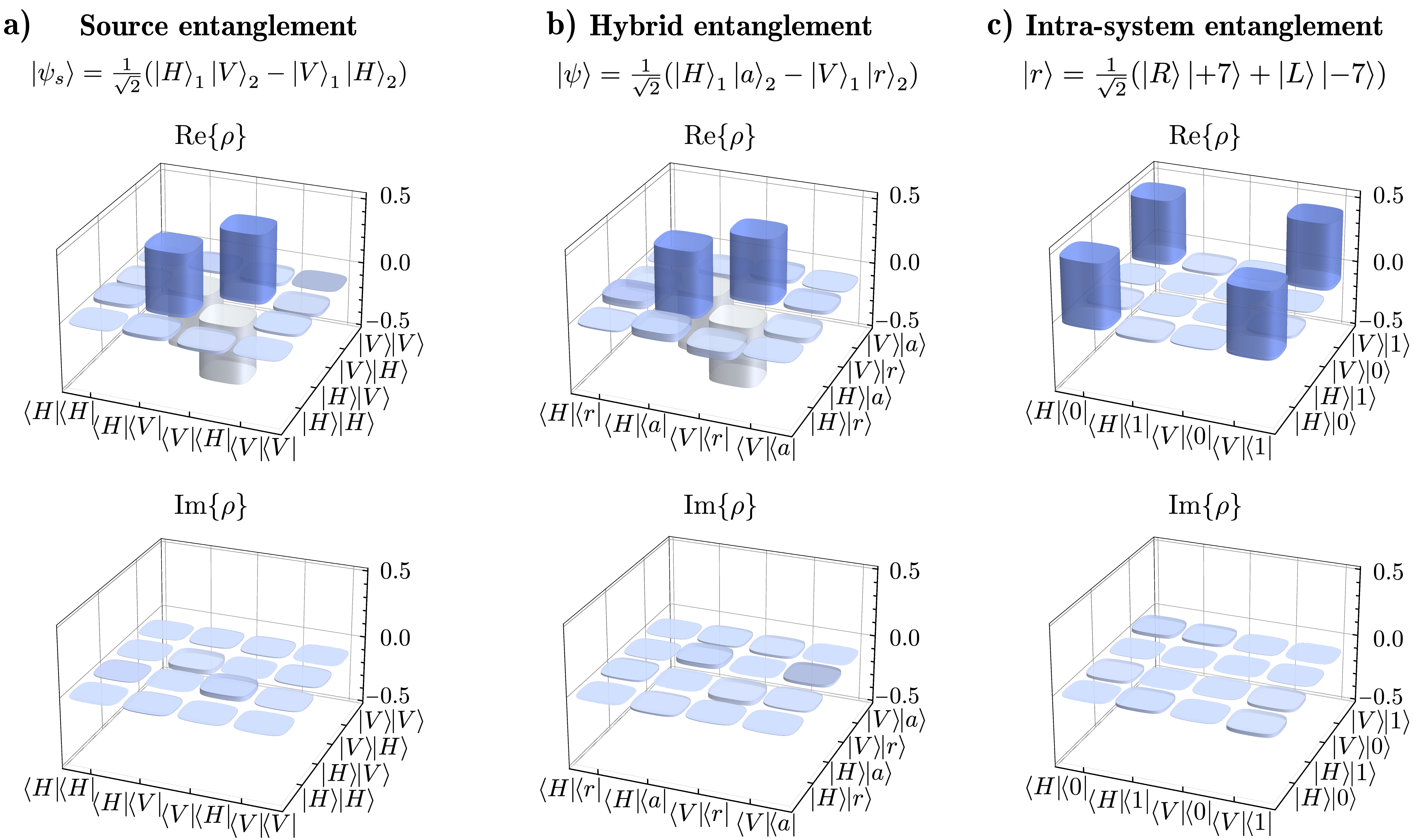}}
    \caption{\textbf{Two-qubit quantum tomographies.} \textbf{a)} Real (top) and imaginary (bottom) parts of the measured density matrix of the polarization entangled state generated by the source, before conversion in OAM.. \textbf{b)} Real (top) and imaginary (bottom) parts of the measured density matrix of the two-photon VV-polarization entangled state after the transmission of photon 2 through the OAM fiber. \textbf{c)} Real (top) and imaginary (bottom) parts of the measured density matrix of the VV  state on photon 2, transmitted through the OAM fiber. The OAM states $\ket{0}$ and $\ket{1}$ in the tomography are defined by the relations: $\ket{0}\equiv (\ket{+7}+\ket{-7})/\sqrt{2}$ and $\ket{1}\equiv i(\ket{-7}-\ket{+7})/\sqrt{2}$. Real and imaginary parts of the experimental density matrices are reconstructed via quantum state tomograpies.}
\label{fig:tomo}
\end{figure*}
In particular, a VV beam belongs to a Hilbert space spanned by  states $\lbrace \ket{R,m}, \ket{L,-m} \rbrace$ \cite{Milione}. For instance, if $m=1$ and if we consider equally distributed superpositions $\ket{r_1}=(\ket{R,+1}+\ket{L,-1})/\sqrt{2}$ and $\ket{a_1}=(\ket{R,+1}-\ket{L,-1})/\sqrt{2}$,  radially and azimuthally polarized beams are obtained \cite{ZhanVV,Cardano}.\\
Here we experimentally demonstrate fiber distribution of a VV-polarization entangled photon state. The conceptual scheme of the experiment is reported in Fig.~\ref{fig:cnpt}. A polarization-VV beam entangled photon pair is generated from an initial polarization entangled pair  (blue ribbon). Only one photon of the pair encodes the VV state (green ribbon). Subsequently, the VV beam is coupled and transmitted through the 5m long air-core fiber and then measured together with the linearly polarized photon. In our case, VV beams of order $m=7$ are generated, whose expressions are:
 \begin{align}
  &\ket{r_7}=\frac{\ket{R,+7}+\ket{L,-7}}{\sqrt{2}}\;   \label{VV7+}\\
  &\ket{a_7}=\frac{\ket{R,+7}-\ket{L,-7}}{\sqrt{2}}
 \label{VV7-}\; .
 \end{align}
The polarization patterns associated to states $\ket{r_7}$ and $\ket{a_7}$ are shown in Fig.~\ref{fig:cnpt}a. In the following, we will refer to $\ket{r_7}$ and $\ket{a_7}$ as $\ket{r}$ and $\ket{a}$, respectively. The experimental apparatus is reported in Fig.\ref{fig:setup}. Polarization entangled photon pairs, $(\ket{H}\ket{V} + \mathrm{e}^{i\phi}\ket{V}\ket{H})/\sqrt{2}$, are generated at 1550 nm wavelength by a periodically poled potassium titanyl phosphate (ppKTP) crystal placed into a polarization Sagnac interferometer and pumped with a continuous-wave laser at 775 nm. Indeed, for fiber-based quantum communication, it is important to exploit photons within the C-band (1530-1565 nm), where optical fibers show minimal losses. The relative phase $\phi$ of the entangled state is  controlled to generate the singlet state:
\begin{equation}
\ket{\psi}_s=\frac{1}{\sqrt{2}}(\ket{H}_1\ket{V}_2 - \ket{V}_1\ket{H}_2)  
\label{s_ent}
\end{equation}
where the subscripts  1 and 2 indicate the two interferometer output modes. According to the notation in Fig. \ref{fig:setup}, photons along output mode 2 impinge on a vortex plate (VP), adding OAM order $m=7$. The VP can be considered as a non-tunable \textit{q}-plate (QP), \textit{i.e.} a device that couples the polarization and the OAM of a single photon \cite{qplate}. Specifically, in the circular polarization basis $\{\ket{R},\ket{L}\}$, the action of a QP with topological charge $q$, on a single photon of OAM order $k$, is described by the transformation: $|L,k\rangle\stackrel{\text{QP}}{\longrightarrow} |R,k+2q\rangle$ and $|R,k\rangle\stackrel{\text{QP}}{\longrightarrow} |L,k-2q\rangle$. In our case, $m=2q=7$ and the initial OAM order is $k=0$. Exploiting the spin-orbit coupling, we generate the VV state $\ket{r}$ or $\ket{a}$ depending on the input polarization state, $\ket{H}$ or $\ket{V}$ respectively \cite{spinorbit1,spinorbit2}. Thus, the polarization entangled singlet state is transformed into the hybrid entangled state:
\begin{equation}
\ket{\psi}=\frac{1}{\sqrt{2}}(\ket{H}_1\ket{a}_2 - \ket{V}_1\ket{r}_2)\;.
\label{hybent}
\end{equation}
 The versatility of our experimental approach is based on full control of each degree of freedom through suitable optical components, allowing the preparation of the desired hybrid VV-polarization entangled state (Eq.(\ref{hybent})).

\subsection*{Hybrid entanglement distribution and measurement}
The main purpose of our work is to prove the feasibility of entanglement distribution with OAM states transmitted through an air-core fiber. In particular, we demonstrate that the coherence of the complex hybrid entangled state in Eq.(\ref{hybent}) is preserved. This is possible since the states $\ket{r}$ and $\ket{a}$ are superposition of anti-aligned states, \textit{i.e.} spin (polarization) and OAM with opposite sign, hence the VV beams are degenerate in time along the fiber transmission \cite{Gregg2015}. Indeed, an eventual non-degeneracy of those states could impair the coherence of the VV states and, therefore, of the entangled pair. The quality of the transmitted states is measured through tomography processes \cite{james2005measurement} and their  entanglement is certified through CHSH-like inequality violations \cite{clauser1969} and tripartite entanglement tests \cite{mermin1990,ardehali1992bell,belinskiui1993,hardy1992,hardy1993,jiang2018hardy}.\\
\textit{Source state.} As a first step, we characterize the initial polarization entangled state in \eqref{s_ent}. To fully determine the quality of the state generated by the ppKTP source, we perform a quantum state tomography within the polarization space of the two photons. The measurements are implemented by collecting two-fold detection after two polarization analysis stages placed along each output mode of the source. 
The obtained tomography is shown in Fig.\ref{fig:tomo}a, where the relative fidelity with respect to the ideal singlet state is $F_{s}=(93.5 \pm 0.2)\%$. Furthermore, we carry out a non-locality test obtaining as the maximum value $S$ of the CHSH inequality $S^{(raw)}_s=2.67\pm0.01$ \cite{clauser1969}. Subtracting the accidental coincidences from $S^{(raw)}_s$, such parameter becomes $S_s=2.68\pm0.01$.\\
\textit{Hybrid entangled state (HyEnt).} Subsequently, we consider the global hybrid VV-polarization entangled state in \eqref{hybent} and measure the two-fold detection after the VV state propagation through the air-core fiber. The fiber structure allows the transmission of OAM modes with very low mode mixing among them. It is composed by a central air core surrounded by a high refractive index ring, creating a large refractive index step that shapes the field of the modes, allowing for their guidance. The fiber we used supports OAM modes with $\ell=\pm5,\pm6,\pm7$ and has 1 dB/km losses \cite{Gregg2015}. In our experiment we have decided to work with modes $\ell=\pm7$, achieving a coupling efficiency $\eta=0.5$. 
At this stage, we consider that the entangled two qubit state lies in a 4-dimensional space spanned by the basis $\{\ket{H}_1\ket{a}_2,\,\ket{V}_1\ket{a}_2,\,\ket{H}_1\ket{r}_2,\, \ket{V}_1\ket{r}_2 \}$, that is composed by the polarization of photon 1 and the VV states of photon 2. The qubit encoded in photon 1 is measured by a polarization analysis stage (green platform in Fig.\ref{fig:setup}). Conversely, the measurements of the VV qubit, \textit{i.e.} photon 2, are implemented by a VP, identical to the one used in the generation process, and a polarization analysis stage (purple platform in Fig.\ref{fig:setup}).
\begin{table}[ht]
\caption{\textbf{CHSH violations.} The CHSH violation parameters obtained from raw data ($S^{raw}$) and by subtracting for accidental coincidences ($S$), are reported for the polarization entangled state generated by the source, the hybrid VV-polarization entangled state (HyEnt) and the intra-system entangled VV state embedded in the photon 2 and transmitted through the air-core fiber (Intra).}
 \centering
\vspace{2mm}
\begin{tabular}{p{1cm} p{2cm} p{2cm} p{2cm}}
\hline
\hline
\rule{0pt}{3ex}State& Measurement Time &$\;\quad S^{(raw)}$&\quad\quad $S$\\[1ex]
\hline
\rule{0pt}{3ex}Source& 160s&$2.67 \pm 0.01$&$2.68 \pm 0.01$\\[1ex]
\rule{0pt}{3ex}HyEnt& 2560s&$2.62 \pm 0.03$&$2.67 \pm 0.03$\\[1ex]
\rule{0pt}{3ex}Intra& 1920s&$2.76 \pm 0.05$&$2.82 \pm 0.05$\\[1ex]
\hline
\end{tabular}
\label{tab:res}
\end{table}
The VP converts back the VV beam to the fundamental Gaussian-like mode, restoring the initial polarization state for photon 2 before impinging on the first VP. In this way, the VV states $\ket{r}$ and $\ket{a}$ are directly mapped into polarization states $\ket{H}$ and $\ket{V}$, which are measured with the usual set of quarter-wave plate (QWP), half-wave plate (HWP) and a polarization beam splitter (PBS) (see Fig.\ref{fig:setup}) \cite{VVcomm2,spinorbit1,spinorbit2}. State tomography in the 4-dimensional space is reported in Fig.\ref{fig:tomo}b. We consider as target state the ideal evolution of the density matrix describing the experimental state generated by the source. The resulting fidelity between such state and the state measured after the fiber propagation is $F_h=(97.9 \pm 0.2)\%$. Furthermore, we observe violation of the CHSH inequality, obtaining the value $S^{(raw)}_h=2.62\pm0.03$ for raw data and value $S_h=2.67\pm0.03$ by subtracting for accidental coincidences, thus violating by 21 and 22 standard deviations the separable limit $S=2$ respectively.\\ 
\begin{figure}[hb!]
   \centering
    {\includegraphics[scale=.3]{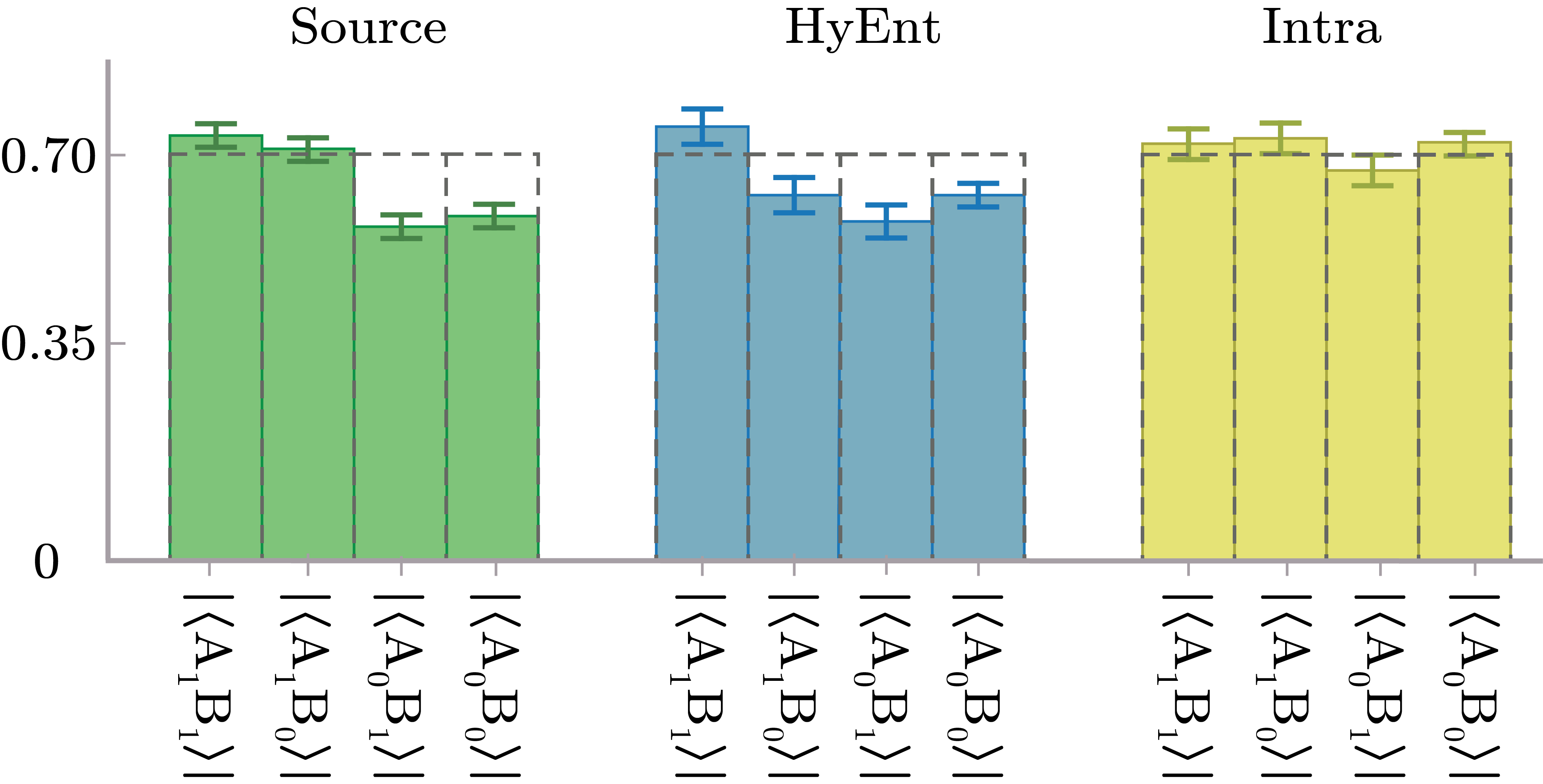}}
    \caption{\textbf{CHSH measurement operators.} Expectation values moduli of the measured operators that maximize the violation of the CHSH parameter $S=\braket{A_1 B_1} - \braket{A_1 B_0} + \braket{A_0 B_1} + \braket{A_0 B_0}$. The values are relative to the polarization entangled state generated by the source (green bars), the hybrid VV-polarization entangled state (blue bars) and the intra-system entangled VV state embedded in the photon 2 and transmitted through the air-core fiber (yellow bars). All error bars are due to Poissonian statistics of the measured events.}
\label{fig:bellop}
\end{figure}

 \begin{figure*}[ht!]
   \centering
    {\includegraphics[scale=.43]{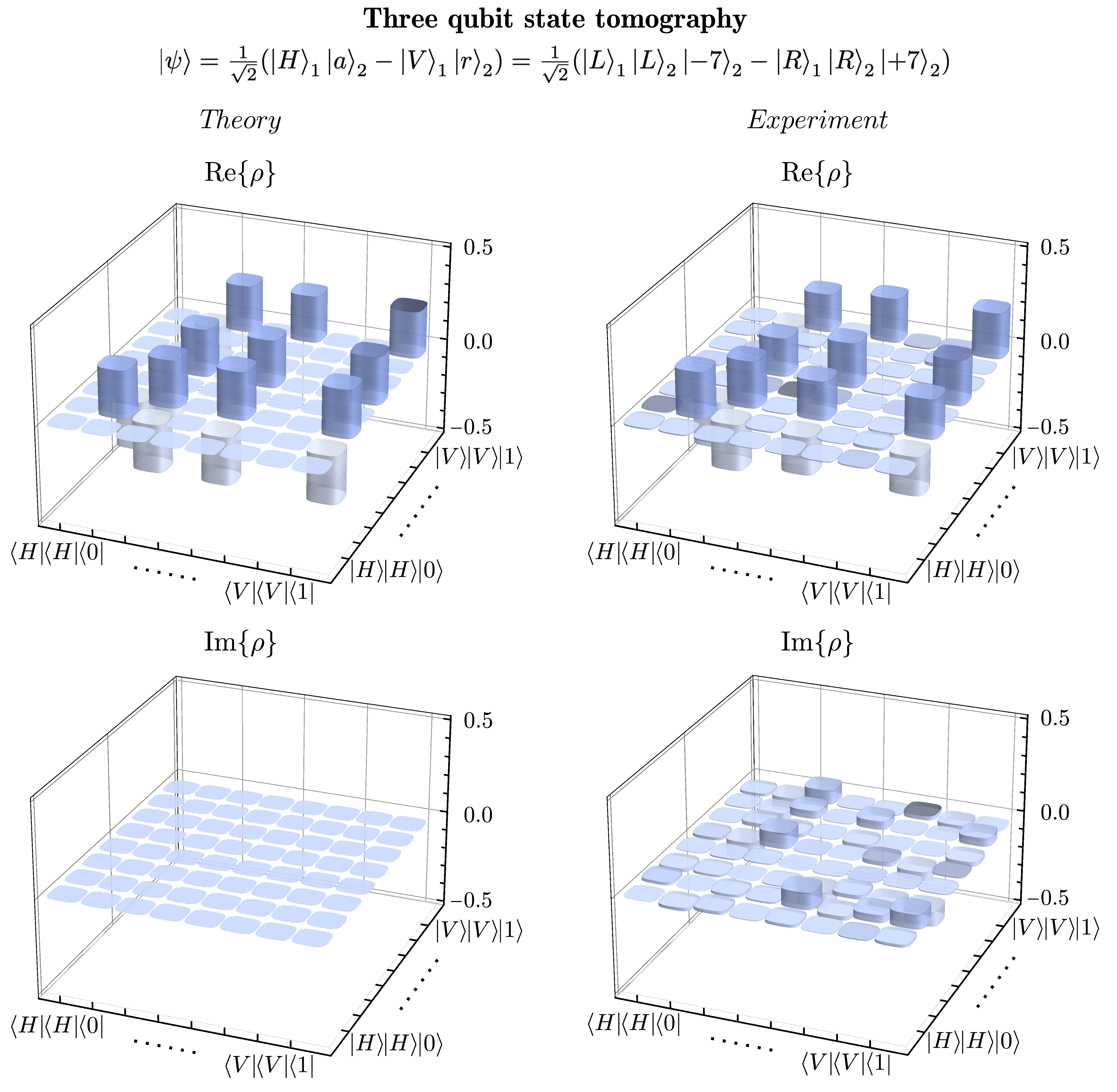}}
    \caption{\textbf{Three-qubit quantum tomography.}  Real and imaginary parts of the measured density matrix  of the hybrid VV-polarization state in space $\{\ket{pol}_1\ket{pol}_2\ket{oam}_2\}$ after the fiber transmission (right) and of the theoretical density matrix of state in \eqref{hybent} (left). The OAM states $\ket{0}$ and $\ket{1}$ in the tomography are defined by the relations: $\ket{0}\equiv (\ket{+7}+\ket{-7})/\sqrt{2}$ and $\ket{1}\equiv i(\ket{-7}-\ket{+7})/\sqrt{2}$. Real and imaginary parts of the experimental density matrices are reconstructed via  quantum state tomography.}
\label{fig:tomo3}
\end{figure*}
\textit{Intra-system entangled state (Intra).} Now, we focus on the VV state embedded in photon 2 and its transmission through the air-core fiber. Such analysis quantifies the quality of the VV beam state generation, transmission through the air-core fiber and conversion to the fundamental Gaussian mode. The single photon VV states $\ket{r}$ and $\ket{a}$, \eqref{VV7+} and \eqref{VV7-}, are maximally entangled in the OAM and polarization degrees of freedom. They corresponds to single-particle entanglement states, referred to as \textit{intra-system} entanglement. The non-separability between polarization and OAM states is not related to nonlocal properties, since they are relative to the same physical system. However, Bell-like inequalities can be exploited to demonstrate the single-particle entanglement, ruling out models that assume realism and non-contexuality of commuting observables, relative to such systems \cite{kocspeck,karimi2010,aiello2015,mclaren2015}. Hence, we certify the presence of intra-system entanglement carrying out quantum state tomography and performing CHSH-like inequality in the space of polarization and OAM degrees of freedom of photon 2. 
Horizontally polarized heralded single photons are sent to the VP to conditionally prepare state state $\ket{r}$ for photon 2. The measurements on the polarization and the OAM degrees of freedom of photon 2 are performed independently. For this purpose, two cascaded measurement stages are needed  (green and purple platforms in Fig.\ref{fig:setup}). A first stage (HWP, QWP and PBS) performs the polarization analysis (green platform  in Fig.\ref{fig:setup}). The second stage composed of a VP and a polarization analysis (purple platform in Fig.\ref{fig:setup}), measures the photon state in the OAM space. As before, the VP maps the information encoded in OAM to a polarization state, which is then measured. Finally, before detection the photon is coupled to a single mode fiber, tracing out all OAM contributions different from the zero order.
The measured quantum state tomography is shown in Fig. \ref{fig:tomo}c and the relative fidelity calculated with respect to the Bell state $\ket{\Phi^+}$ is $F_i=(99.4 \pm 0.6)\%$. The corresponding parameters $S_i$ obtained from the CHSH-like inequality violations are $S_i^{(raw)}=2.76\pm0.05$ and $S_i=2.82\pm0.05$. The set of CHSH violations measured for each state (source, HyEnt and Intra) is summarized in table \ref{tab:res} and the mean values of the measured operators are shown in Fig.\ref{fig:bellop}. \\
\textit{Three qubits HyEnt.} The previous measurements have independently certified the high fidelity of both the hybrid VV-polarization entangled state and the single photon VV beam state after propagation in the air-core fiber.
To complete the characterization of the hybrid VV-polarization entangled state in \eqref{hybent}, we now measure such state with a different apparatus that does not assume a 2-dimensional Hilbert space for photon 2 spanned by $\{\ket{r},\ket{a}\}$. In this case, the state in \eqref{hybent} corresponds to a 3-qubit state that lives in the $2^3$ dimensional Hilbert space spanned by the basis of three qubits $\{\ket{pol}_1\ket{pol}_2\ket{oam}_2\}$, encoded in the polarizations of the two photons and the OAM degree of freedom of photon 2.
The measurements on the final state  are performed by a polarization analysis stage for photon 1, and by polarization and OAM analysis stages for photon 2, as the one adopted for the intra-system entanglement characterization (Fig.\ref{fig:setup}). This apparatus allows to measure independently the polarization and the OAM component of photon 2. Thus, performing the $2^3$-dimensional quantum state tomography of the transmitted state (Fig.\ref{fig:tomo3}), a final fidelity $F=(88.1 \pm 0.2)\%$ with respect to the ideal state in Eq.(\ref{hybent}) is obtained, thus showing that the fiber preserves the injected state. As for the other cases, also for the 3-qubit case we perform a device independent test of the quantum correlations, showing their preservation after fiber transmission of the VV state. First, we test the Mermin-Ardehali-Belinski\v{i}-Klyshko inequality \cite{mermin1990,ardehali1992bell,belinskiui1993}, which provides an upper bound for contextual hidden variable theories describing the correlations between observables relative to three qubits:
\begin{equation}
\begin{split}
 \mathcal{M}\quad\equiv \quad |\braket{A_1 B_2 C_2}&+\braket{A_2 B_1 C_2}+\\
 &+\braket{A_2 B_2 C_1}-\braket{A_1 B_1 C_1}|\leq2 \; .
\label{Mermino}
\end{split}
\end{equation}
The observables $A_i$, $B_i$ and $C_i$ ($i=1,2$) are dichotomic (with eigenvalues $\pm1$) and relative to the first, the second and the third qubit, respectively. Violation of such inequality certifies the nonclassical correlations of tripartite states. Furthermore, if a value $\mathcal{M}\ge 2\sqrt{2}$ is found, models in which quantum correlations are allowed between just two of the three qubits (biseparable quantum models), are ruled out as well \cite{collins2002,bancal2011}. The state \eqref{hybent} in the $2^3$-dimensional space is able to reach the algebraic value of $\mathcal{M}=4$ by choosing the operators: $A_1=-\sigma_z^A$, $A_2=\sigma_x^A$,  $B_1=-\sigma_z^B$, $B_2=\sigma_x^B$, where $\sigma_i$ ($i=x,z$) are the Pauli operators relative to photons 1 (A) and 2 (B) in the polarization in basis $\{\ket{H},\ket{V}\}$; and $C_1=\sigma_z^C$, $C_2=\sigma_x^C$, where the Pauli operators are in the OAM basis $\{\ket{0}\equiv (\ket{+7}+\ket{-7})/\sqrt{2}, \,\;\ket{1}\equiv i(\ket{-7}-\ket{+7})/\sqrt{2}\}$ relative to photon 2. Measuring such operators after the VV state transmission and calculating the parameter $\mathcal{M}$, we obtain $\mathcal{M}^{(raw)}=3.43\pm0.04$ from raw data, and the value $\mathcal{M}=3.53\pm0.04$ by subtracting accidental coincidences. In this way, we violated the classical bound by 35 and 38  standard deviations and the quantum biseparable bound by 15 and 17 standard deviations, respectively.\\
Finally, we further study the correlation of the state in \eqref{hybent} by performing a Hardy test \cite{hardy1992,hardy1993},  recently generalized  in a suitable form for more than two parties by \cite{jiang2018hardy}. Given a system with certain null correlation probabilities, a paradox arises when other events are automatically forbidden in the framework of noncontextual hidden variable models, while they can happen within a quantum context. 
Since experimentally measuring null probabilities represents a difficult task, Hardy logical contradictions can be conveniently mapped into more general inequalities. In Ref. \cite{jiang2018hardy} an extended multi-party version of Hardy's paradox is proposed, leading to an inequality that for three qubits reads:
\begin{equation}\begin{split}
\mathcal{H}\quad\equiv &\quad P(A_1A_2A_3)-P(A_1B_2B_3)-P(A_1\bar{B}_2\bar{B}_3)+\\
 &-P(B_1A_2B_3)-P(\bar{B}_1A_2\bar{B}_3)-P(B_1B_2A_3)+\\
 &-P(\bar{B}_1\bar{B}_2A_3)\le0 \; ,
 \end{split}
\label{difficile}
\end{equation}
where $A_i$ ($B_i$) represents a dichotomic operator $A$ ($B$) acting on qubit $i=1,2,3$ with eigenvalues $\pm1$, $\bar{B_i}\equiv -B$ and the probabilities $P(X_1Y_2Z_3)\equiv P(X_1=1,Y_2=1,Z_3=1)$.
In our case, the transmitted 3-qubit state permits to maximally violate the generalized Hardy test by choosing the operators: $A_1=A_2=-A_3=\sigma_z$ and $B_1=B_2=B_3=\sigma_x$ relative to the qubits $\ket{pol}_1$ and $\ket{pol}_2$ (in basis $\{\ket{H},\ket{V}\}$) and $\ket{oam}_2$ (in basis $\{\ket{0},\ket{1}\}$), respectively.
The experimental value $\mathcal{H}$ obtained for raw data is $ \mathcal{H}^{(raw)}=0.085\pm0.008$ and by accounting for accidental coincidences it becomes $\mathcal{H}=0.104\pm0.008$ (theoretical value for the ideal state is $\mathcal{H}=0.25$). Such values allow to violate the noncontextual bound by 10 and 12 standard deviations, respectively. Note that, the tripartite correlations obtained are generated by both contextual (intrasystem) and nonlocal (intersystem) entanglement.

\subsection*{Conclusions and discussion}
Future quantum communication will require to distribute quantum states over long distances.. The protocols implemented within such systems will include the distribution of high-dimensional and entangled quantum states. Indeed, spanning Hilbert spaces of greater dimensions allows higher information capacity and noise resilience, leading to enhanced quantum information processing \cite{Erhard2018,bechmann2000,cerf2002}. In this context, VV states represent a powerful resource for classical and quantum applications.\\
Here, we demonstrated the feasibility of distributing complex VV states through an OAM supporting fiber, also permitting to preserve entanglement with a different system. To fully assess the robustness to decoherence and quality of the transmitted complex entangled state, we performed quantum state tomographies, violations of CHSH-like inequalities and multipartite entanglement tests. The achieved  fidelities of the transmitted state demonstrate the capability to perform high fidelity distribution in an OAM supporting fiber of a hybrid VV-polarization entangled state at the telecom wavelength. In particular, the possibility to simultaneously encode and distribute information in the polarization and OAM degree of freedom of a single particle represents an useful resource due to the higher robustness to losses, while tools for their processing have been identified \cite{vitelli2013}. This work paves the way towards adoption of high-dimensional entanglement in quantum networks. Further perspectives of this work involve the investigation of fiber-based distribution of different orders of OAM entangled states and their distribution over longer distances, exploiting the potential scalability arising from a fiber-based approach. Other perspectives involve interfacing of OAM integrated circuits \cite{cai2012integrated,oamchip,chiptofiber} through OAM supporting fibers for future quantum networks.

\vspace{10pt}
\textit{Note added}. During the preparation of this manuscript, the authors became aware of a work by Huan Cao \textit{et al.} on a similar topic \cite{cina}.

\section*{Funding Information}
This work is supported by the Center of Excellence, SPOC - Silicon Photonics for Optical Communications (ref DNRF123), by the People Programme (Marie Curie Actions) of the European Union's Seventh Framework Programme (FP7/2007-2013) under REA grant agreement n$^\circ$ $609405$ (COFUNDPostdocDTU), and by the ERC-Advanced grant PHOSPhOR (Photonics of Spin-Orbit Optical Phenomena; Grant Agreement No. 694683). G.C acknowledges Becas Chile and Conicyt.

\section*{Acknowledgments}
We thank S. Ramachandran and P. Gregg for the fiber design, P. Kristensen from OFS-Fitel for the fiber fabrication and D. Poderini for many  advices on the software development.

\end{document}